\documentclass{webofc}
\usepackage[varg]{txfonts}   
\usepackage{algorithm}
\usepackage{enumerate}
\usepackage{algpseudocode}
\usepackage{mathptmx}
\usepackage{amsmath}
\usepackage{amssymb}
\usepackage{amsfonts}
\usepackage{graphicx}
\usepackage{subfigure}

\begin{document}
\title{Efficient and Scalable Approach to Equilibrium Conditional\\ Simulation of Gibbs Markov Random Fields}
%
%

\author{\firstname{Milan} \lastname{\v{Z}ukovi\v{c}}\inst{1}\fnsep\thanks{\email{milan.zukovic@upjs.sk}} \and
        \firstname{Dionissios T.} \lastname{Hristopulos}\inst{2}\fnsep\thanks{\email{dionisi@mred.tuc.gr}}
}

\institute{Institute of Physics, Faculty of Science, P. J. \v{S}af\'arik University, Park Angelinum 9, 041 54 Ko\v{s}ice, Slovakia
\and
Geostatistics Laboratory, Technical University of Crete, Chania 73100, Greece
          }

\abstract{%
We study the performance of an automated hybrid Monte Carlo (HMC) approach for conditional simulation of a recently proposed, single-parameter Gibbs Markov random field (Gibbs MRF).  The MRF is based on a modified version of the planar rotator (MPR) model and is used for efficient gap filling in gridded data. HMC combines the deterministic over-relaxation method and the stochastic Metropolis update with dynamically adjusted restriction and performs automatic detection of the crossover to the targeted equilibrium state. We focus on the  ability of the algorithm to efficiently drive the system to equilibrium at very low temperatures even with sparse conditioning data. These conditions are the most challenging computationally, requiring extremely long relaxation times if simulated by means of the standard Metropolis algorithm. We demonstrate that HMC has considerable benefits  in terms of both computational efficiency and prediction performance of the MPR method.
}
\maketitle
\section{Introduction}
\label{intro}
Gaussian Markov random fields (GMRFs)  are used for modeling spatial data on regular grids~\cite{Rue05}. GMRFs are based on the principle of conditional independence and the enforcement of spatial correlations via local interactions. The latter translate into sparse precision matrices, which allow computationally efficient representations. While GMRFs have a  long history~\cite{Gelfand10}, non-Gaussian Markov random fields (NGMRFs) have attracted less attention. Binary-valued Ising spin models, the $q$-state Potts model, and the continuous planar rotator are typical NGMRF examples. Widely studied in statistical physics, they have also found  applications in areas such as image restoration~\cite{nishi99,tadaki01,saika02} and spatial prediction~\cite{mz-dth09}.

Recently, we have introduced a novel Gibbs Markov random field for prediction of spatial data on regular grids, based on the modified planar rotator (MPR) model~\cite{mz-dth18}. We also proposed an efficient and automated hybrid Monte Carlo (HMC) approach for the conditional simulation of the model. HMC has been shown to lead to fast relaxation, and the short-range nature of the interaction between the ``spin'' variables enabled vectorization. Consequently, the MPR computational time for both inference and simulation was found to scale approximately linearly with system size, which makes MPR-based prediction attractive for big and gappy data sets, such as satellite and radar images.  
Given a rectangular grid of size $L_x \times L_y$, the problem of interest is to estimate by efficient updating the unknown values  at $P$ prediction sites (missing data), while the conditioning values (sample data) at the remaining sites are kept fixed during the updating process.

Herein we focus on the HMC approach and study its performance via a vis the standard Metropolis algorithm. The latter is known to be inefficient at very low temperatures, which is the operating parameter region of the MPR method. We show that HMC updates can considerably reduce the relaxation times of the standard Metropolis approach; even more importantly, the number of HMC sweeps necessary to reach equilibrium is insensitive to grid size, i.e., the HMC algorithm is scalable.

\section{Hybrid Monte Carlo}
\label{hmc}

\begin{algorithm}[t!]
\caption{HMC algorithm. $\boldsymbol{\hat{\Phi}}^{\mathrm{old}}$ is the
current and $\boldsymbol{\hat{\Phi}}^{\mathrm{new}}$ is the new spin state. $\boldsymbol{\hat{\Phi}}^{\mathrm{old}}_{-p}$ is the current state excluding the point labeled by $p$. $U(0,1)$ denotes  the uniform probability distribution in $(0, 1)$.}
\label{algo:mpr-relax}
\begin{algorithmic}
\Procedure{HMC}{$\boldsymbol{\hat{\Phi}}^{\mathrm{new}},\boldsymbol{\hat{\Phi}}^{\mathrm{old}},a,T$}
\For{$p=1, \ldots, P$}  \Comment Loop over prediction sites
\State 1: ${\hat{\Phi}'}_p \gets \mathcal{O}\{\hat{\Phi}_p^{\mathrm{old}} \}$ \Comment Perform \emph{over-relaxation update} according to~\eqref{eq:over-relax}
\State 2:   $r_1 \gets U(0,1)$ \Comment Generate uniform random number
\State 3:   ${\hat{\Phi}''}_p \gets {\hat{\Phi}'}_p + 2\pi (r_1 -0.5)/a \pmod{2\pi}$ \Comment Propose spin update
\State 4:   $ \Delta \mathcal{H} = \mathcal{H}({\hat{\Phi}''}_p, \boldsymbol{\hat{\Phi}}^{\mathrm{old}}_{-p}) - \mathcal{H}({\hat{\Phi}'}_p, \boldsymbol{\hat{\Phi}}^{\mathrm{old}}_{-p})$ \Comment Calculate energy change
\State 5:  $W = \min\{1,\exp(-\Delta \mathcal{H}/T)\}$ \Comment Calculate acceptance probability
\State 6:  $\boldsymbol{\hat{\Phi}}^{\mathrm{new}}_{-p} \gets \boldsymbol{\hat{\Phi}}^{\mathrm{old}}_{-p}$ \Comment Perform \emph{Metropolis update}
            \If{$W > r_2 \gets U(0,1)$}
            \State 6.1: {$\hat{\Phi}_p^{\mathrm{new}} \gets {\hat{\Phi}''}_p$}    \Comment Accept the new state
            \Else
            \State 6.2: {$\hat{\Phi}_p^{\mathrm{new}} \gets {\hat{\Phi}'}_p$} \Comment Keep the current state
\EndIf
\EndFor \Comment End of prediction loop
\State 7: \Return $\boldsymbol{\hat{\Phi}}^{\mathrm{new}}$  \Comment Return the updated state after one HMC sweep
\EndProcedure

\end{algorithmic}
\end{algorithm}

The standard Metropolis algorithm is often used in MC simulation due to its flexibility and applicability to a wide range of problems~\cite{metro53}. However, it can be rather inefficient in some situations, e.g. at low temperatures, due to very low acceptance rate (proportional to $\exp(-\Delta E/T)$, where $\Delta E = E^{\mathrm{new}}-E^{\mathrm{old}}$ is the energy difference between the new and old states). This leads to extremely long relaxation times in the low-$T$ limit, which is the typical parameter region for the MPR prediction $(T \approx 10^{-2})$~\cite{mz-dth18}. Efficient use of the MPR method requires an updating scheme that is able to drive the system to equilibrium fast, i.e., with the shortest possible relaxation time.

To tackle this problem we proposed the HMC updating scheme (see Algorithm~\ref{algo:mpr-relax}). HMC combines a \emph{flexible restricted form of stochastic Metropolis} and the \emph{deterministic over-relaxation}~\cite{creutz87} methods. The former algorithm generates a proposal spin-angle state at the \emph{i}th site according to the rule $\phi_i'=\phi_i+2\pi(r-0.5)/a$, where $r$ is a uniformly distributed random number $r \in (0,1)$. A tunable parameter $a$ is automatically reset during the equilibration to maintain the acceptance rate above a predefined threshold value (arbitrarily set to $0.3$). In the over-relaxation update, a new spin-angle value at the \emph{i}th site is chosen so that the system energy  is conserved. In the MPR model defined by the nearest–neighbor interaction Hamiltonian ${\mathcal H}=-J\sum_{\langle i,j \rangle}\cos[(\phi_i-\phi_j)/2]$, the over-relaxation update is achieved by means of the following transformation
\begin{equation}
\label{eq:over-relax}
\phi'_{i}   = \, \left[2\, \arctan2 \left( \sum_{j \in nn(i)} \sin{\phi_{j}}, \sum_{j \in nn(i)} \cos{\phi_{j}} \right)-\phi_{i}\right] \mod {2\pi},
\end{equation}
where $nn(i)$ denote the nearest neighbors of $\{ \phi_{i}\}_{i=1}^{P}$, and $\arctan2(\cdot)$ is the four-quadrant inverse tangent: for any real $x, y$ such that $|x| + |y| >0$, $\arctan2(y, x)$ is the angle (in radians) between the positive horizontal axis and the point  $(x, y)$.

\section{Results and Conclusion}
\label{res}

\begin{figure}[t!]
\centering
\subfigure{\label{fig:N_relax-L_T001_p09}\includegraphics[width=7cm,clip]{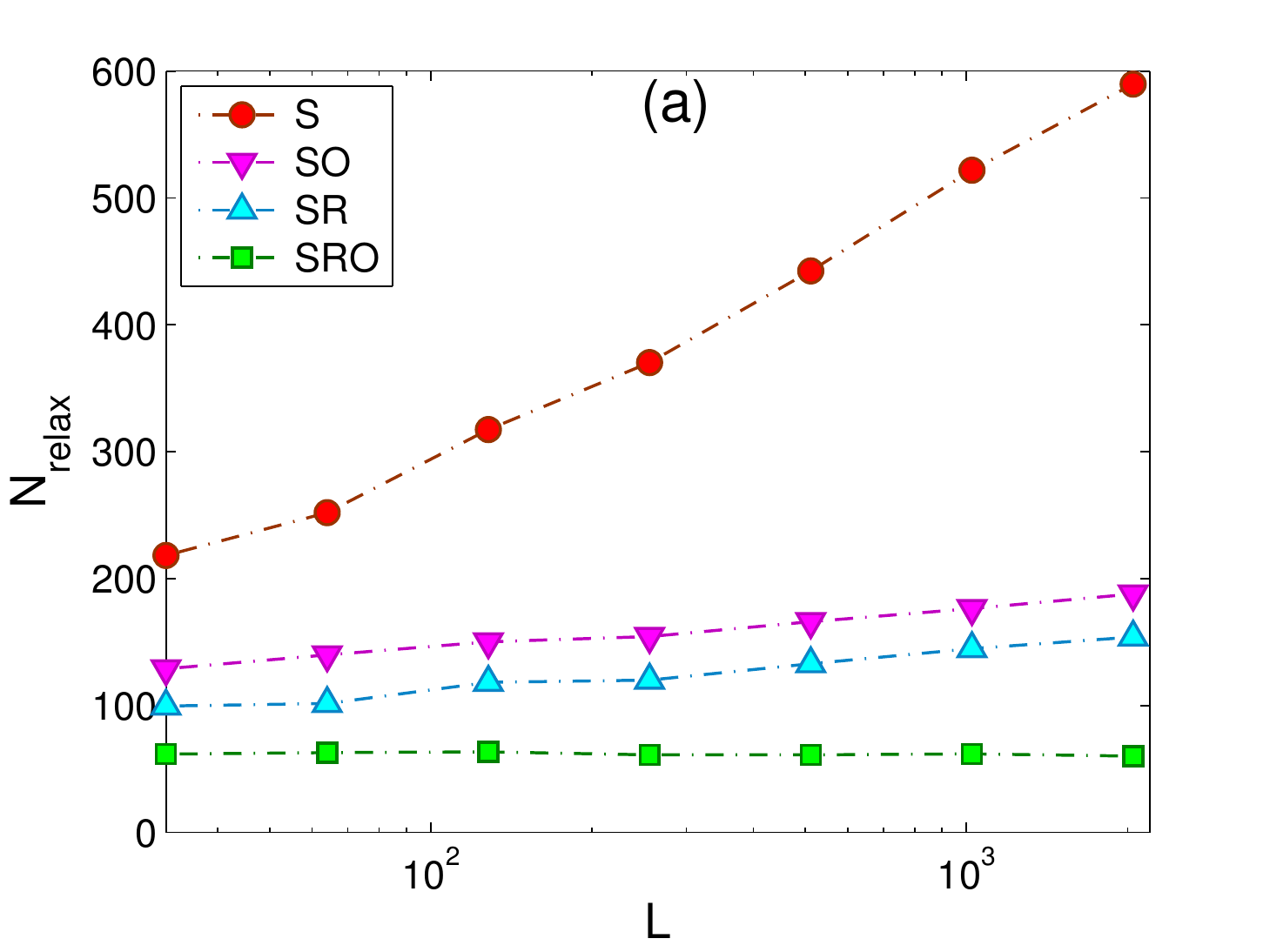}}
\subfigure{\label{fig:e-N-L_T001_p09}\includegraphics[width=7cm,clip]{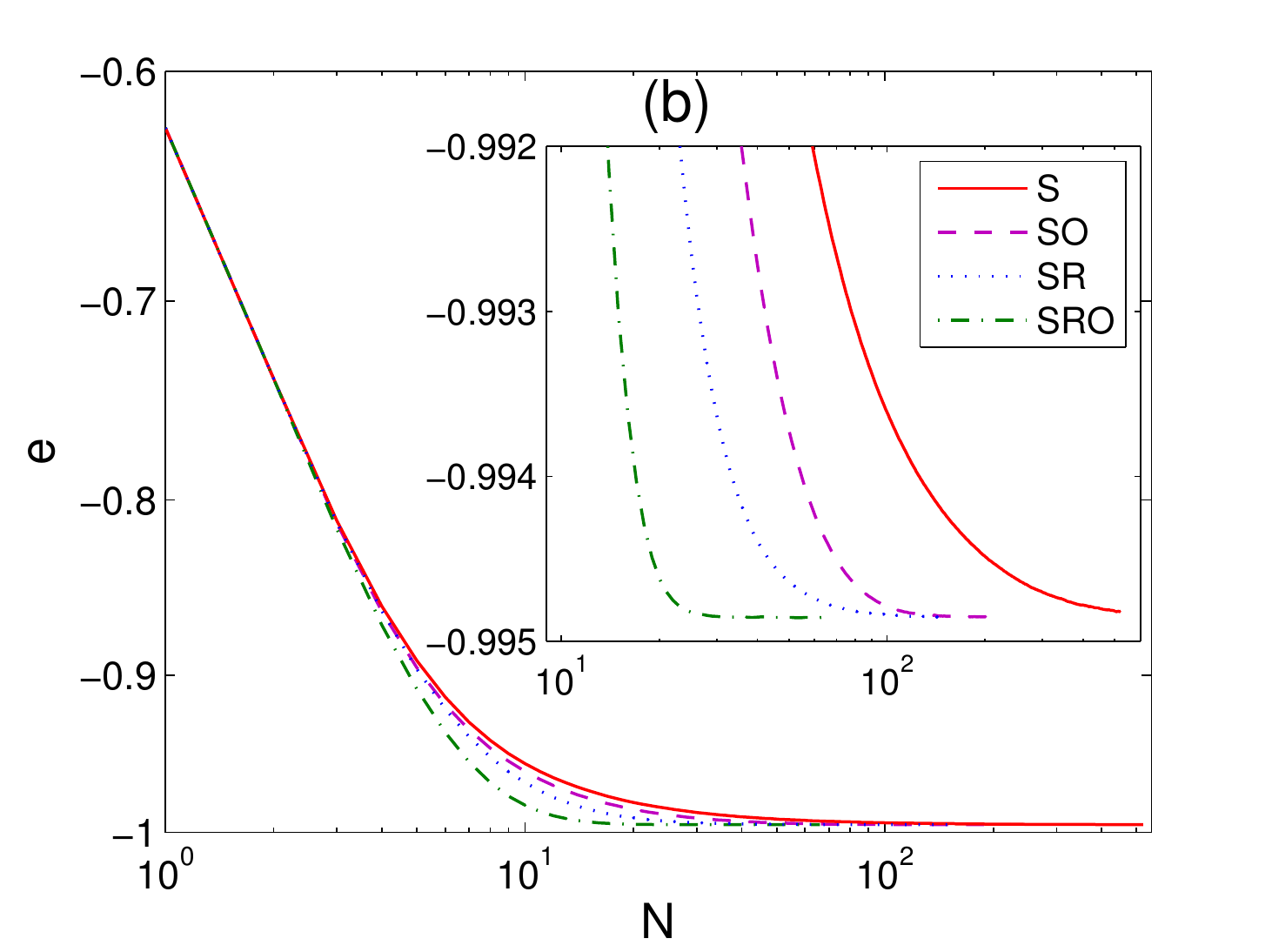}}
\caption{(a) $N_{\mathrm{relax}}$ as a function of $L$, for $c=90\%$ and $T=0.01$, obtained by the standard (S), standard and over-relaxed (SO), standard and restricted (SR) and hybrid, i.e., standard, restricted and over-relaxed (SRO) Metropolis algorithms. (b) Evolution of the specific energy $e=\langle {\mathcal H} \rangle/L^2$ to equilibrium, for $L=2048$.}
\label{fig:N_relax-L_M_2fix}       
\end{figure}

Efficiency of the HMC approach is demonstrated on Gaussian synthetic data  $Z \sim N(m=50,\sigma=10)$ and exponential covariance $C(r) =\sigma^{2} \exp(-r/\xi)$ where $\xi=5$, simulated on square grids with $L$ nodes per side  $(L=32,\hdots, 2048)$. We simulate missing data by randomly removing $c\%$ of the $L^2$ values. Ensemble expectations  are obtained
by averaging over different sampling configurations.  The relaxation time is expressed as the number of MC sweeps, $N_{\mathrm{relax}}$, necessary to reach equilibrium.

\begin{figure}[t!]
\centering
\subfigure{\label{fig:e_min-L_T001_p09}\includegraphics[width=7cm,clip]{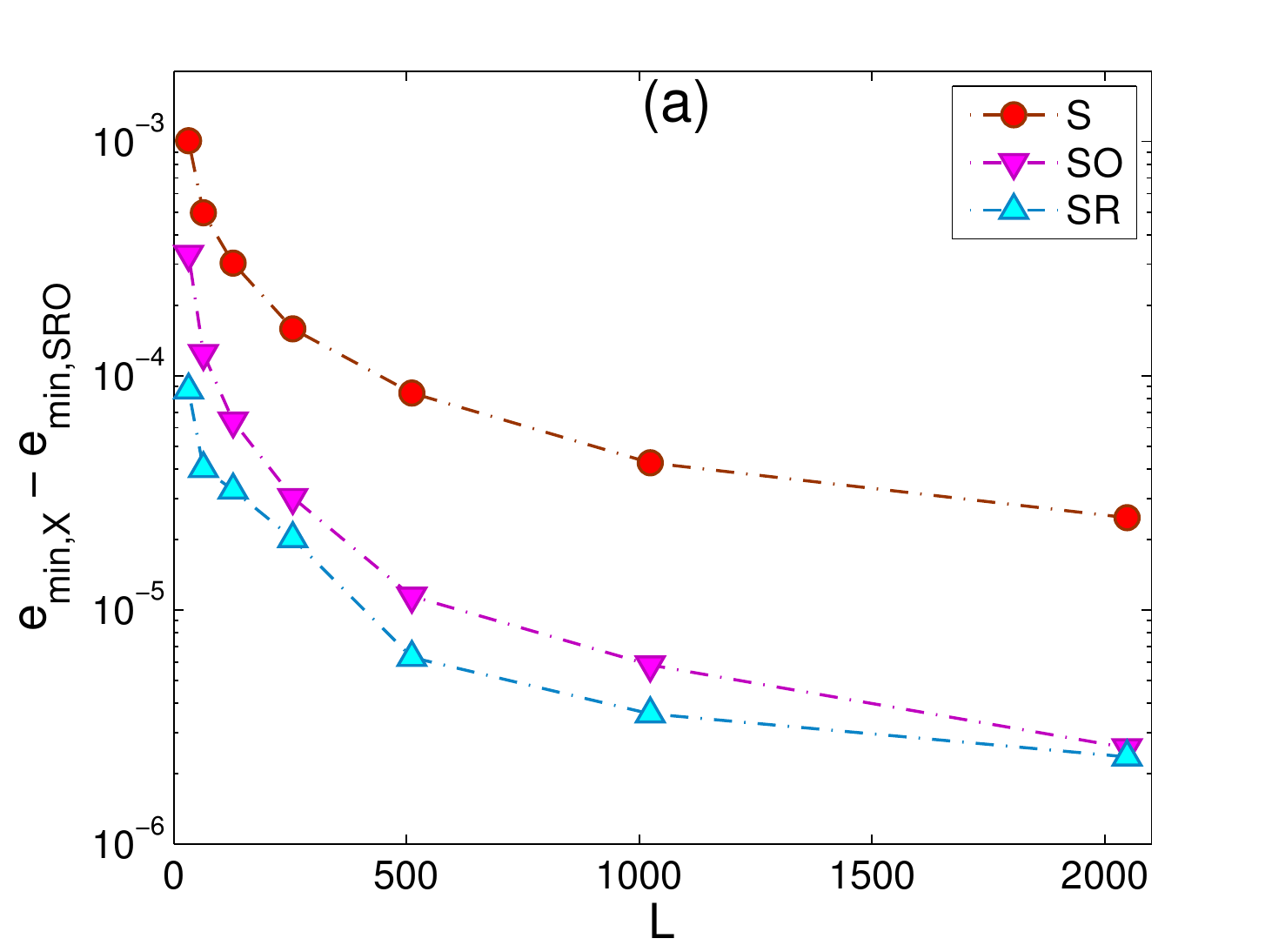}}
\subfigure{\label{fig:rmse-L_T001_p09}\includegraphics[width=7cm,clip]{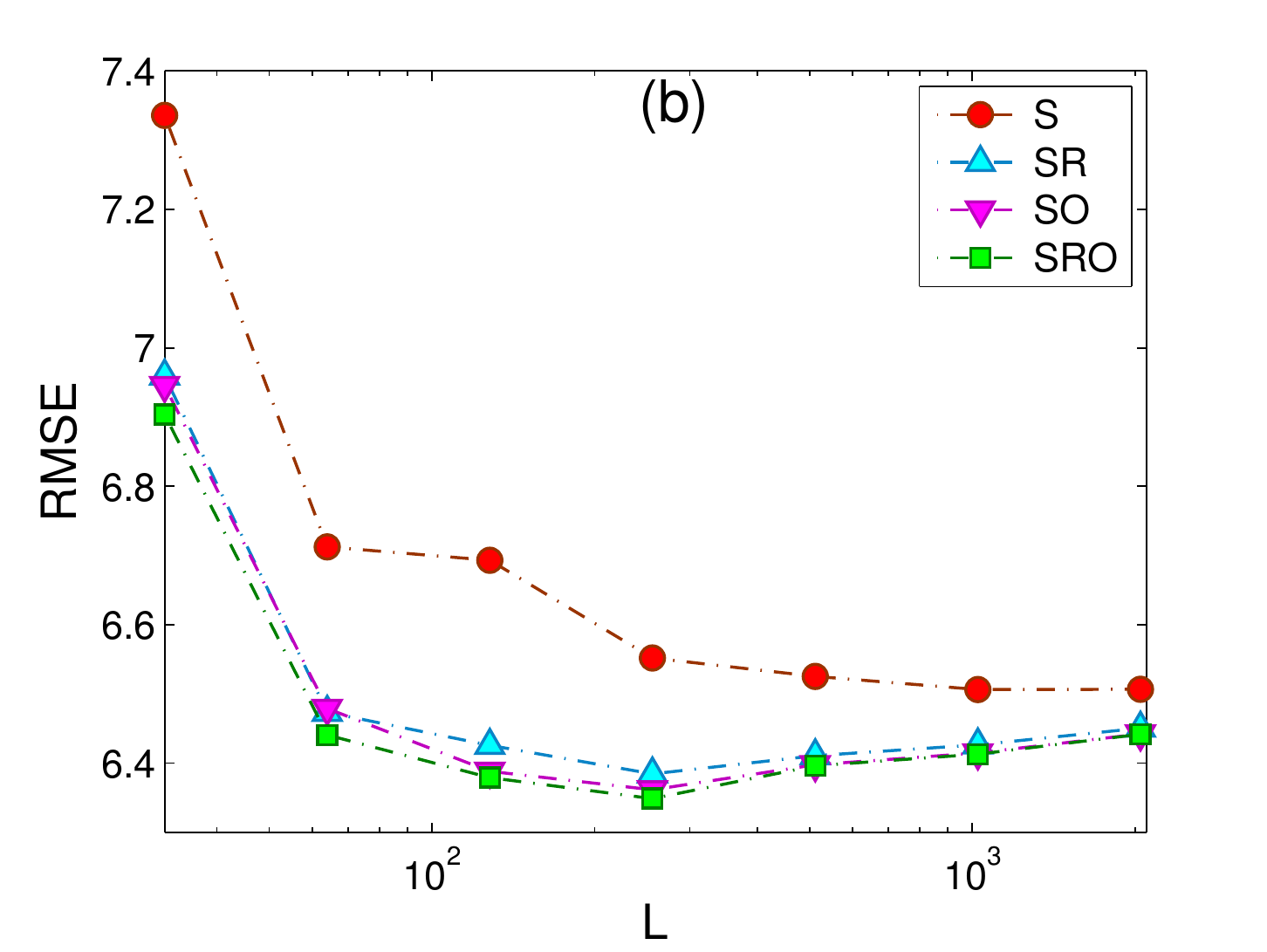}}
\caption{(a) Differences in the specific energy minima achieved at the equilibrium onset $N_{\mathrm{relax}}$, using the X (= S, SO and SR) and the hybrid (SRO) algorithms. (b) RMSE error curves for different algorithms.}
\label{fig:e_min}       
\end{figure}

We focus on $N_{\mathrm{relax}}$ in the limit of sparse samples and low temperatures. As shown in Figure~\ref{fig:e_min-L_T001_p09}, for $c=90\%$ and $T=0.01$, both $N_{\mathrm{relax}}$ and its slope with increasing $L$ are  largest for standard Metropolis updating. On the other hand, $N_{\mathrm{relax}}$ and its rate of increase are considerably suppressed by combining standard Metropolis with either over-relaxation or restricted updating. However, the hybrid method that combines all three approaches  further suppresses $N_{\mathrm{relax}}$ significantly and completely eliminates its dependence on $L$. Thus, relaxation by merely $\approx 60$ hybrid MC sweeps suffices to equilibrate the sparsely conditioned system (for $L$ ranging between 32 and 2048). Figure~\ref{fig:e-N-L_T001_p09} shows the evolution of the specific energy for $L=2048$. Relaxation from the random initial state to the equilibrium (flat) regime is detected automatically at the crossover point $N=N_{\mathrm{relax}}$, where the trend disappears. The inset shows that the standard Metropolis update converges to equilibrium  much slower than the hybrid method; in addition, standard Metropolis fails to reach a perfectly flat regime even at $N_{\mathrm{relax}}$. This could be addressed by a stricter convergence criterion, which would further increase the relaxation times. Therefore, albeit large,  $N_{\mathrm{relax}}$ for standard Metropolis is in fact an underestimate.

The differences in minimum energy values (reached at $N_{\mathrm{relax}}$) between simpler methods and the hybrid approach are illustrated in Figure~\ref{fig:e_min-L_T001_p09}. Figure~\ref{fig:rmse-L_T001_p09} shows the root mean square error ${\rm RMSE} =\sqrt{\sum_{\vec{s}_{p} \in G_{p}}\left[Z(\vec{s}_p) - \hat{Z}(\vec{s}_p)\right]^2/P}$, where $Z(\vec{s}_p)$ represent true and $\hat{Z}(\vec{s}_p)$ predictions at the missing points $\vec{s}_p, \ p=1,\hdots,P$. Due to slower relaxation to equilibrium, MPR gap filling with either standard (S) or only partially hybrid (SO or SR) updating schemes yields  larger RMSEs than the hybrid method.

In summary, we demonstrated that the HMC algorithm can significantly increase both the computational and prediction performance in the challenging limits of very low temperature and sparse conditioning data. Moreover, the HMC relaxation time is insensitive to grid size, i.e., the algorithm is scalable. Owing to the short-range nature of the interactions between variables, the computational efficiency of the HMC algorithm (and the entire MPR method), can be further increased by vectorization or parallelization on graphics processing units~\cite{mz_etal08}. This advance will make MPR gap filling attractive for near real-time processing of big data sets, e.g. satellite and radar images.

\end{document}